\documentclass[journal]{IEEEtran}
\usepackage{amsmath,amsfonts}

\usepackage{algorithm}
\usepackage{bm}
\usepackage{array}
\usepackage{mathtools}
\usepackage{color,soul}
\usepackage{siunitx}

\def\BibTeX{{\rm B\kern-.05em{\sc i\kern-.025em b}\kern-.08em
T\kern-.1667em\lower.7ex\hbox{E}\kern-.125emX}}
\usepackage{balance}
\usepackage{multirow}
\usepackage{cite}

\usepackage[utf8]{inputenc}

\usepackage{breqn}
\usepackage[inkscapeformat=png]{svg}
\usepackage{stackengine}

\DeclareMathOperator{\Var}{{Var}}
\DeclareMathOperator{\Cov}{{Cov}}
\DeclareMathOperator{\sinc}{sinc}

\DeclareMathOperator{\E}{\mathbb{E}}

\DeclareMathOperator{\En}{\mathcal{E}} 

\DeclareMathOperator{\T}{\text{T}} 

\makeatletter
\newcommand{\@giventhatstar}[2]{\left[#1\;\middle|\;#2\right]}
\newcommand{\@giventhatnostar}[3][]{#1[#2\;#1|\;#3#1]}
\newcommand{\giventhat}{\@ifstar\@giventhatstar\@giventhatnostar}
\newcommand{\Fourier}{\mbox{\boldmath$\cal F$}}

\makeatother

\begin{document}
\title{Filtered Multi-Tone Spread Spectrum with Overlapping Subbands}
\author{Brian Nelson, Hussein Moradi, and Behrouz Farhang-Boroujeny}

\maketitle

\begin{abstract}
  A new form of the filter bank multi-carrier spread spectrum (FBMC-SS) waveform is presented. This new waveform modifies the filtered multi-tone spread spectrum (FMT-SS) system, and is intended to whiten the power spectral density (PSD) of the transmit signal. In the conventional FMT-SS, subcarrier bands are non-overlapping, leaving a spectral null between the adjacent subcarrier bands. To make FMT-SS more appealing for a broader set of applications than those studied in the past, we propose adding additional subcarriers centered at these nulls and thoroughly explore the impact of the added subcarriers on the system performance. This modified form of FMT-SS is referred to  as overlapped FMT-SS (OFMT-SS). We explore the conditions required for maximally flattening the PSD of the synthesized OFMT-SS signal and for cancelling the interference caused by overlapping subbands. We also explore the choices of spreading gains that result in a low peak-to-average power ratio (PAPR) for a number of different scenarios. Further reduction of the PAPR of the synthesized signal through clipping methods is also explored. Additionally, we propose methods of multi-coding for increasing the data rate of the OFMT-SS waveform, while minimally impacting its PAPR.
\end{abstract}

\copyright 2025 IEEE. Personal use of this material is permitted. Permission from IEEE must be obtained for all other uses, in any current or future media, including reprinting/republishing this material for advertising or promotional purposes, creating new collective works, for resale or redistribution to servers or lists, or reuse of any copyrighted component of this work in other works.

\section{Introduction}
The use of filter bank multi-carrier spread spectrum (FBMC-SS) has been reported in a number of publications, e.g., \cite{HaraS1997OomC,KondoS1996PomD,HanzoLajosL2005OoMC, DarylUCC, HaabDavidB.2021SSSD, Taylor_NMF}. All these works make use of filtered multi-tone (FMT) for signal synthesis, thus, may be referred to as FMT-SS. As compared to the more conventional spread spectrum techniques, such as, direct sequence spread spectrum (DS-SS) and frequency hopping spread spectrum (FH-SS) methods, FMT-SS is shown to be more robust to partial-band interferers, as the filter bank structure allows straightforward removal of interferences at a cost of reduced processing gain, \cite{DarylUCC}.

\begin{figure}
  \centering
  \includegraphics[width=0.95\columnwidth]{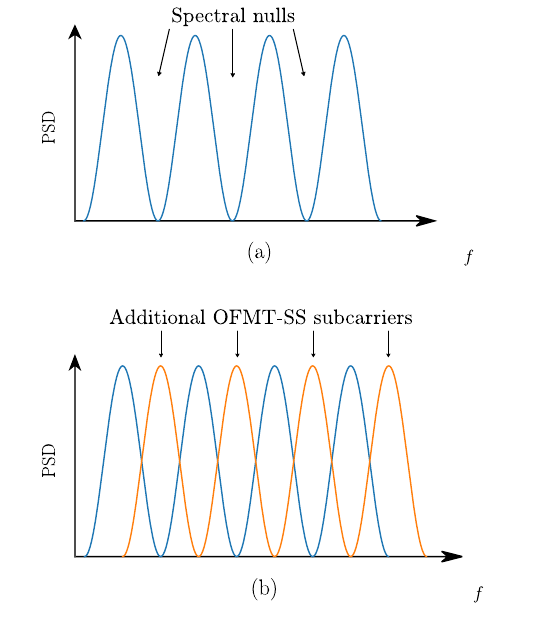}
  \caption{A plot of subcarrier PSDs in (a) FMT-SS and (b) OFMT-SS. In OFMT-SS, the additional subcarriers, indicated in orange color, are added to fill in the null bands. }
  \label{fig:fill_gaps_fmt}
\end{figure}

Though FMT-SS has the above advantages when compared to conventional spread spectrum techniques, it is known to suffer from a non-flat power spectral density (PSD) which follows since subcarrier bands, by design, are non-overlapping. This leaves a spectral null between each pair of adjacent subcarrier bands, i.e. a portion of the transmission band without any transmission power. An example of such PSD is presented in Fig.~\ref{fig:fill_gaps_fmt}(a). For a given transmit power, the presence of spectral nulls in the PSD of the synthesized signal increases the peak(s) of the PSD, hence, results in an increased interference to other communications over the active subcarrier bands. In addition, the presence of these spectral nulls reveals a transmitter's identity and parameters to an unauthorized observer that wishes to detect the transmitted information and/or to intercept the communication.

For many applications, having a flat transmit PSD can provide other benefits as well. Here, we highlight three well-suited applications to FBMC-SS that would benefit from a flat transmission PSD: (i) underlay communications for network setup in a cognitive radio system, (2) skywave high-frequency (HF) for long-haul communications, (3) ultra-wideband (UWB) communications over short ranges.

In \cite{DarylUCC, 7503836}, FMT-SS was suggested as an underlay communication channel to exchange spectral information among the nodes in a cognitive radio network. To minimize interference to other radios that may be using the same spectrum, it will be advantageous to keep the PSD of the synthesized signal as low as possible.

FMT-SS application for communications over skywave HF channels (in the 3 to 30 MHz spectrum) has been explored in the past \cite{LarawayStephenAndrew2020PAoa, HaabDavidB2018FBMS}. In this domain, spectral allocations are nearly uncontrolled and thus interference among different users of the HF spectrum are likely. Keeping the PSD of the transmitted signals flat, hence, minimizing the peak of the PSD, here also is of great value.

Ultra-wideband (UWB) waveforms have been proposed for short range communications over the spectral bands that have been licensed for a variety of applications. In order to avoid interference with the licensed users of these spectral bands, the Federal Communications Commission (FCC) has mandated that the PSD of any UWB waveform remains below a spectral mask at the transmitter antenna output \cite{HeydariP.2005Asol}. The use of filter banks for UWB communications have been recently proposed in \cite{UWB2024}. Making use of a form of FBMC-SS with a flat PSD will double the transmission power compared to its FMT-SS counterpart, hence, may significantly improve on the transmission range.

To remove the FMT-SS spectral profile, an alternative approach that results in a flat spectrum was proposed in \cite{austin_smt_channel_estimation}. In this work, FMT is replaced by the staggered multi-tone (SMT) modulation, \cite{OFDMvsFBMC}. SMT uses offset quadrature amplitude modulation (OQAM) for overlapping of adjacent subcarrier bands, hence, leading to a near flat PSD of the synthesized signal. However, the design presented in \cite{austin_smt_channel_estimation} uses an FMT-SS preamble to perform synchronization and channel estimation, thus maintaining the undesirable spectral profile for a portion of the transmission.

In this paper, we introduce an alternative FBMC-SS approach with a flat PSD that can be used for both the preamble and payload portions of the packet. This approach modifies the FMT-SS approach discussed in \cite{DarylUCC} by inserting additional subcarriers over the null bands with the goal of flattening the PSD of the synthesized signal. The modified waveform is referred to as  overlapped FMT-SS (OFMT-SS). An example of the PSD of subcarriers of OFMT-SS is presented in Fig.~\ref{fig:fill_gaps_fmt}(b).

Although inserting subcarriers in the FMT null bands removes regular null bands from its spectral profile, overlapping of the subcarriers may introduce other problems including inter-carrier interference (ICI) and some other form of fluctuations in the PSD of the synthesized signal. We introduce design methods that minimize the ICI as well as PSD fluctuation caused by overlapping of subcarriers. We also present design approaches that keep the peak-to-average power ratio (PAPR) of the synthesized signal at a minimum level, while increasing data rate through a multi-coding method.

To be more specific, this paper makes the following contributions.
\begin{itemize}
  \item
        OFMT-SS waveform is proposed as an effective multi-carrier spread spectrum method.
  \item
        A pulse-shape design that removes ICI among adjacent subcarriers and leads to a perfectly flat spectrum in the transmit signal is developed.
  \item
        A multi-code method that enables variable data rate OFMT-SS transmissions is explored. Making use of simulated annealing, a code design method that results in a relatively low PAPR is presented.
  \item
        Methods that limit the PAPR of OFMT-SS signals are suggested. These methods are shown to be very effective in this application. It is shown that the PAPR of OFMT-SS can be reduced to as low as $4$~dB (or, even lower) with a negligible error rate performance loss.
\end{itemize}

The rest of this paper is organized as follows. Section~\ref{sec:FMT} presents a summary of FMT-SS. Here, the basic results of FMT-SS that are relevant to the remaining parts of this paper are summarized. Section~\ref{sec:ofmt_intro} introduces the OFMT-SS waveform and explores how its pulse shape can be constrained/designed to remove ICI and perfectly flatten the PSD of the transmit signal. Section~\ref{sec:ofmt_papr} presents a strategy for designing the OFMT-SS pulse shape to minimize PAPR while the constraints outlined in Section~\ref{sec:ofmt_intro} are imposed.  A design of multi-codes for increasing data rate, while minimizing PAPR is also presented. In Section~\ref{sec:papr_reduction}, we explore the application of clipping methods to further reduce the PAPR of OFMT-SS. The performance of the design is confirmed through bit-error rate (BER) and symbol-error rate (SER) simulations in Section~\ref{sec:simulations}. The concluding remarks are made in Section~\ref{sec:conclusion}.

\section{FMT-SS Summary}\label{sec:FMT}

In FMT-SS, spreading chips are across a set of non-overlapping subcarrier bands. Accordingly, the transmit pulse shape/filter is given as
\begin{equation} \label{eq:fmt-ps}
  g(t) = \sum_{k=0}^{N - 1} \gamma_k h(t)e^{j 2 \pi f_k t}
\end{equation}
where $N$ is the number of subcarriers, the coefficients $\gamma_k$ are a set of complex-valued and unit-amplitude spreading gains, $h(t)$ is the prototype filter, and $f_k$ is the center frequency of the $k^{th}$ subcarrier band.

In \cite{DarylUCC}, it is shown that the combined impulse response of the transmit and receive filters, i.e., $\eta(t) = g(t) \star g^*(-t)$, can be factored as
\begin{equation} \label{eq:fmt_impulse_response}
  \eta(t) = \beta(t)\rho(t)
\end{equation}
where
\begin{equation} \label{eq:fmt_rho}
  \rho(t) = h(t) \star h^*(-t)
\end{equation}
and
\begin{equation} \label{eq:fmt_beta}
  \beta(t) = \sum_{k=0}^{N - 1} e^{j2\pi f_k t}
\end{equation}
Moreover, by design, $h(t)$ is a real-valued and symmetric square-root Nyquist filter. Accordingly, $\rho(t)$ is a real-valued and symmetric Nyquist filter.

\begin{figure}
  \centering
  \includegraphics[width=0.95\columnwidth]{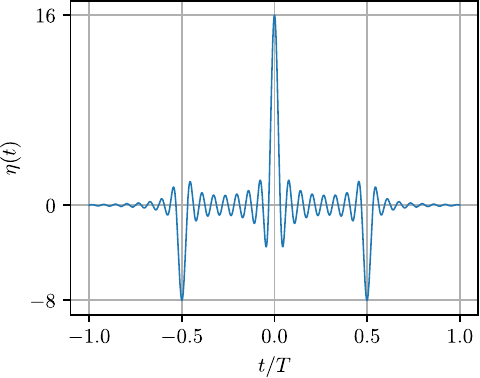}
  \caption{An example plot of $\eta(t)$ in FMT-SS for the case where $N=16$.}
  \label{fig:eta(t)}
\end{figure}

In the FMT-SS design of \cite{DarylUCC}, the subcarriers are spaced at $2/T$ Hz apart, where $T$ is the symbol interval. This allows a prototype filter design with maximum transition bandwidth, thus minimizing the cost of implementing the underlying filter bank by allowing a lower order filter for the same stopband attenuation. In addition, the subcarrier frequencies, at the baseband, are chosen to be
\begin{equation}\label{eq:fkFMT-SS}
  f_k=\frac{2(k-N/2)+1}{T}, ~~~\mbox{for }k=0, 1, \cdots, N-1.
\end{equation}
It is also shown that with these choices, the combined response $\eta(t)$ consists of three `sinc' pulses at the time positions $t=-T/2$, $0$, and $T/2$. An example of $\eta(t)$ is presented in Fig.~\ref{fig:eta(t)}. Using the preamble structure described in \cite{DarylUCC}, these pulses may be used for carrier acquisition and timing recovery at the receiver.

In certain applications, it is exceedingly important to design the FMT-SS waveform to synthesize signals with a minimum peak-to-average power ratio (PAPR). A particular case that FMT-SS has been applied to and requires a low PAPR is communication over high-frequency (HF) skywave channels, where transmit power can be many tens or, even, many hundreds of watts. Hence, power efficiency of the transmitter power amplifier becomes a key design factor.  PAPR minimization of FMT-SS has been explored in \cite{LarawayStephenAndrew2015Hbfb}. To this end, it has first been noted that \eqref{eq:fmt-ps} can be rearranged as
\begin{equation} \label{eq:fmt-ps2}
  g(t) = h(t)m(t)
\end{equation}
where
\begin{equation}\label{eq:m(t)}
  m(t)=\sum_{k=0}^{N - 1} \gamma_k e^{j 2 \pi f_k t}.
\end{equation}
Next, it has been argued that to avoid large peaks in the synthesized signal (equivalently, to minimize the PAPR), the spreading gain factors $\gamma_k$ have to be chosen such that the crest-factor of
$m(t)$ is minimized. The crest-factor of a signal is defined as its maximum amplitude over its root-mean-square (RMS) value. Friese \cite{Friese_algorithm} has proposed an algorithm for minimizing the crest-factor of $m(t)$. This algorithm leads to  the choices of $\gamma_k$ that results in a crest-factor of as low as $1.07$  and for the synthesized FMT-SS signal a PAPR of as low as $3.61$~dB.

Even though the emphasis above and the study of reducing PAPR of FMT-SS in \cite{LarawayStephenAndrew2015Hbfb} was motivated by its specific application to HF communications, PAPR minimization is of interest in all applications. By minimizing the PAPR, transmitter power amplifiers can operate more efficiently and device power consumption can be reduced. With this point in mind a significant part of this paper is allocated to PAPR reduction in OFMT-SS; see Sections~\ref{sec:ofmt_papr} and \ref{sec:papr_reduction}.

\section{OFMT-SS} \label{sec:ofmt_intro}
While in FMT-SS the subcarrier center frequencies, at the baseband, may be expressed as in \eqref{eq:fkFMT-SS}, for OFMT-SS they are chosen as
\begin{equation}\label{eq:fkOFMT-SS}
  \bar f_k=\frac{k-N+\frac{1}{2}}{T}, ~~~\mbox{for }k=0, 1, \cdots, L-1
\end{equation}
where
\begin{equation} \label{eq:def_L}
  L = 2N.
\end{equation}
Note that if the symbol rate is kept the same in FMT-SS and OFMT-SS, the above choices of $ \bar f_k$ minimally affect the transmission bandwidth of OFMT-SS when compared to its FMT-SS counterpart. Also, to keep the equations distinguishable between FMT-SS and OFMT-SS, we add an over-bar for the variables/functions that are related to OFMT-SS and are different from their counterpart in FMT-SS.
Hence, the OFMT-SS pulse-shaping filter is expressed as
\begin{equation} \label{eq:fs-ofmt-ps}
  \bar{g}(t) = \sum_{k=0}^{L - 1}  \bar\gamma_k h(t)e^{j 2\pi \bar f_k t}.
\end{equation}
Note that both $g(t)$ and $\bar{g}(t)$ are based on the same prototype filter $h(t)$. However, as noted above, to keep the transmission bandwidth the same, the number of subcarrier bands are doubled. Reducing the spacing between subcarriers leads to some differences in the system impulse response.

Here, the combined response of $\bar{g}(t)$ and its matched version leads to the system impulse response
\begin{align} \label{eq:ofmt_impulse_response}
  \bar{\eta}(t) & = \bar{g}(t) \star \bar{g}^*(-t) \nonumber                                \\
                & = \left(\sum_{k=0}^{L - 1} \bar\gamma_k h(t)e^{j 2\pi \bar f_k t} \right)
  \star \left(\sum_{k=0}^{L - 1} \bar\gamma^*_{k} h(t)e^{j 2\pi \bar f_k t} \right).
\end{align}
Taking note that only aligned and adjacent subcarrier bands overlap, \eqref{eq:ofmt_impulse_response} can be rearranged to
\begin{equation} \label{eq:ofmt_impulse_response_simp}
  \begin{aligned}
    \bar{\eta}(t) & = a(t) + c(t)
  \end{aligned}
\end{equation}
where
\begin{align} \label{eq:def_aligned_terms}
  a(t) & = \sum_{k=0}^{L - 1}  \bar\gamma_k h(t)e^{j 2\pi \bar f_k t} \star  \bar\gamma^*_k h(t)e^{j 2\pi \bar f_k t}
\end{align}
accounts for the aligned terms, and
\begin{align} \label{eq:def_cross_terms}
  c(t) & = \sum_{k=0}^{L - 2}  \bar\gamma_k h(t)e^{j 2\pi \bar f_k t} \star  \bar\gamma^*_{k+1} h(t)e^{j 2\pi \bar f_{k+1} t} \nonumber \\
       & \quad + \sum_{k=0}^{L - 2}  \bar\gamma^*_k h(t)e^{j 2\pi \bar f_k t} \star  \bar\gamma_{k+1} h(t)e^{j 2\pi \bar f_{k+1} t}
\end{align}
accounts for the crossed terms between the adjacent subcarrier bands. The remaining terms are ignored, as in a well-designed prototype filter they have a negligible contribution to the result of \eqref{eq:ofmt_impulse_response}. By factoring the impulse response into these two sets of terms, we are able to gain insight into how the reduced subcarrier spacing impacts the shape of the PSD and the ICI. We continue by analyzing $a(t)$ and $c(t)$ separately.

\subsection{Aligned Terms} \label{ssec:FS-FMT-aligned}
\begin{figure}
  \centering
  \includegraphics[width=0.95\columnwidth]{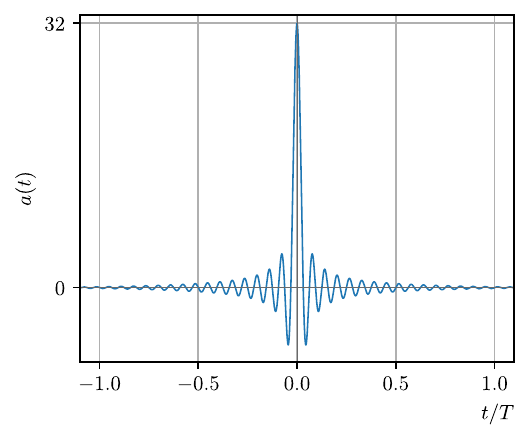}
  \caption{An example plot of $a(t)$ in OFMT-SS for the case where $L=15$.}  \label{fig:bar-eta(t)}
\end{figure}

Straightforward manipulations lead to
\begin{equation} \label{eq:aligned_terms}
  a(t) = \bar{\beta}(t) \rho(t)
\end{equation}
where
\begin{equation} \label{eq:beta_bar_def}
  \bar{\beta}(t) = \sum_{k=0}^{L - 1} e^{j 2\pi \bar f_k t},
\end{equation}
$\rho(t)$ is given by \eqref{eq:fmt_rho}, and $\bar f_k$ is given by \eqref{eq:fkOFMT-SS}. Here, we note that $\bar{\beta}(t)$ has a similar form to $\beta(t)$ defined in equation \eqref{eq:fmt_beta}, but with a frequency spacing of $1/T$ Hz between the sinusoidal terms instead of $2/T$~Hz. By using $1/T$ Hz as the frequency spacing of the sinusoids in \eqref{eq:beta_bar_def}, $\bar{\beta}(t)$ is a $\sinc$ pulse train with a spacing of $T$ between peaks. This leads us to conclude that $a(t)$ is similar to $\eta(t)$ shown in Fig. \ref{fig:eta(t)}, but with the sinc pulses at time instants $\pm T/2$ removed and the height of the $\sinc$ pulse at $t=0$ doubled. An example of $a(t)$ is presented in Fig.~\ref{fig:bar-eta(t)}.

The above results show that $a(t)$ is a Nyquist pulse with regular zero crossings at the  interval $T/L$. This implies that, if one could force $c(t)$ to zero, the pulse-shape $\bar g(t)$ would be a square-root Nyquist pulse with a bandwidth of $L/T$, leading to a {\em flat} spectrum across the band of transmission. An alternative way of looking at this result is to first recall that $h(t)$ is chosen so that $\rho(t)$ is a Nyquist pulse with bandwidth $1/T$ and roll-off factor $\alpha$. Second, since $a(t)$ is the sum of $L$ copies of $\rho(t)$ modulated to integer multiples of $1/T$, $a(t)$ is a Nyquist pulse with bandwidth $L/T$ and roll-off factor $\alpha/L$.  Next, we discuss the constraints that need to be imposed to force $c(t)$ to zero.

\subsection{Cross Terms} \label{ssec:FS-FMT-ICI}
To analyze $c(t)$, we start by defining
\begin{equation} \label{eq:single_ici_term}
  d(t) = h(t) \star h(t)e^{j \frac{2 \pi}{T}t}.
\end{equation}
After applying some straightforward manipulations to \eqref{eq:def_cross_terms}, and making use of \eqref{eq:single_ici_term}, one finds that
\begin{equation} \label{eq:c_simplified}
  c(t) = d(t) \sum_{k=0}^{L - 2} e^{j 2\pi f_k t} \left(\bar\gamma^*_k \bar\gamma_{k+1}  + \bar\gamma_k \bar\gamma^*_{k+1}\right).
\end{equation}

We would like to choose the spreading gains $\bar\gamma_k$ so that
\begin{equation} \label{eq:c(t)=0}
  c(t) = 0.
\end{equation}
By solving this equation, we guarantee that $\bar\eta(t) = a(t)$, hence, the OFMT-SS signal will have  flat spectrum, i.e., a white PSD across the band of transmission.

For \eqref{eq:c(t)=0} to be satisfied, the spreading gains should be chosen so that
\begin{equation} \label{eq:constrain_c_0_all_times}
  \bar\gamma^*_k \bar\gamma_{k+1} + \bar\gamma_k \bar\gamma^*_{k+1} = 0 \quad \mbox{for } k = 0, 1, \ldots, L - 2.
\end{equation}
Defining the spreading gain vector as
\begin{equation} \label{eq:spreading_gain_vector_def}
  \bm{\bar\gamma} = \begin{bmatrix}
    \bar\gamma_0 & \bar\gamma_1 & \dotsb & \bar\gamma_{L - 1}
  \end{bmatrix}^\T,
\end{equation}
it can be shown that the only spreading sequences that satisfy \eqref{eq:constrain_c_0_all_times} have the form:
\begin{equation} \label{eq:no_ici_full_samp_rate_constraint}
  \bm{\bar\gamma} =e^{j\phi} {\bf b} \odot \bm \zeta
\end{equation}
where
\begin{equation} \label{eq:def_b}
  { \bf b} = \begin{bmatrix}
    1 & j & j^2 & \cdots & j^{L-1}
  \end{bmatrix}^\T,
\end{equation}
$\odot$ denotes an element-wise product, $\phi$ is an arbitrary phase, and the elements of $\bm \zeta$ are given by
\begin{equation} \label{eq:real_spr_gains}
  \zeta_k \in \left\{-1, +1 \right\}, \mbox{ for }0\le k\le L-1.
\end{equation}

The result in \eqref{eq:no_ici_full_samp_rate_constraint} can be proved by letting $\bar\gamma_k = e^{j\phi_k}$  and  $\bar\gamma_{k+1} = e^{j\phi_{k + 1}}$ in \eqref{eq:spreading_gain_vector_def} and noting that this leads to
\begin{equation} \label{eq:single_gamma_constraint1}
  \cos(\phi_{k + 1} - \phi_k)  = 0.
\end{equation}
This result implies
\begin{equation} \label{eq:cos_is_0}
  \phi_{k + 1} - \phi_k = n\pi \pm \frac{\pi}{2}.
\end{equation}
Letting $\bar\gamma_0=\pm e^{j\phi}$ and making use of \eqref{eq:cos_is_0}, one can see that \eqref{eq:no_ici_full_samp_rate_constraint} holds.

In the rest of this paper, for simplicity, and without any loss of generality of the optimization methods that will be introduced, we let $\bar\gamma_0=1$.

\section{OFMT-SS PAPR Minimization} \label{sec:ofmt_papr}
PAPR minimization is desirable in all communication systems. Here, we present a design strategy that allows us to select the elements of the vector $\bm \zeta$ with the goal of minimizing the PAPR of the  synthesized OFMT-SS signal. In our previous works on FMT-SS, we found that PAPR minimization of such signals is closely related to minimization of the crest-factor of the multi-tone signal $m(t)$ in \eqref{eq:m(t)}, \cite{LarawayStephenAndrew2015Hbfb}. The same is true here, with the difference that the spreading coefficients $\bar\gamma_k$ are constrained to the discrete choices expressed by \eqref{eq:no_ici_full_samp_rate_constraint}. In FMT-SS the spreading coefficients $\gamma_k$ can choose any unit amplitude complex value. This allows the use of the Friese optimization method for minimizing the crest-factor of $m(t)$, \cite{Friese_algorithm}.

Unfortunately, there is no similar method to Friese optimization that works well when the spreading coefficients are allowed to select values from a finite/discrete set. Although an approach for finding discrete spreading gain coefficients resulting in a moderate crest-factor of approximately 2 has been suggested in \cite{1085837}, it is recognized that these coefficients are likely not optimal. It turns out that the only way of finding the set of $\bar\gamma_k$ that minimizes the crest-factor of
\begin{equation}
  \bar m(t) = \sum_{k=0}^{L-1} \bar\gamma_k e^{j2\pi f_k t}
\end{equation}
is to search over all $2^L$ choices of the vector  $\bm \zeta$. This clearly has a complexity that grows exponentially with $L$ and, thus, is not scalable as $L$ grows. Common methods of solving this type of problems make use of statistical search approaches, such as, Markov chain Monte Carlo (MCMC), genetic, or simulated annealing algorithms. Here, we have chosen to use a simulated annealing algorithm, following \cite{DENGH1996Sobs}.

Simulated annealing is a stochastic optimization algorithm where a cost function is minimized in a manner similar to how the molecules in materials cool to lower energy states. In the simulated annealing algorithm, an initial state is chosen at random and used to evaluate the cost function. At each iteration of the algorithm, the state is perturbed, and the cost of the new state is measured. If the cost is reduced, the perturbed state is accepted. If the cost increases, then the new state is accepted with some probability that depends on the increase in cost and on the ``temperature'' of the algorithm. This process of perturbation and state updates is repeated until an ``equilibrium'' is reached. We follow the definition of equilibrium used in \cite{DENGH1996Sobs} by using the standard deviation of the cost at the current temperature to detect equilibrium. When the algorithm converges to an equilibrium state, the temperature is lowered and the algorithm is continued at the new temperature. Lowering the temperature reduces the probability of accepting a state that increases the cost. The algorithm exits once the cost of the state vector has not changed for several steps of the temperature. By gradually reducing the temperature, the algorithm converges to a global optimum. Simulated annealing was proposed in \cite{doi:10.1126/science.220.4598.671} and has been discussed in many publications, in particular for optimization over a discrete space. Some examples include \cite{Cerny1985Thermodynamical,LaarhovenP.J.M.vanPeterJ.M.1987Sa:t,DENGH1996Sobs}.

As the starting point of our simulated annealing algorithm, we follow the approach in \cite{DENGH1996Sobs}. Here, the author has explored using simulated annealing to design binary sequences with good aperiodic autocorrelation and cross-correlation properties. The good autocorrelation here is quantified by the relative size of the autocorrelation coefficients for non-zero lags when compared to the signal power (i.e., autocorrelation with a zero lag). The sequences with smaller non-zero lag autocorrelation coefficients are considered as better sequences. It turns out that sequences with small non-zero lag autocorrelation coefficients result in a lower crest-factor when such sequences are used to construct a multi-tone signal like $\bar m(t)$; see \cite{Friese_algorithm}. Because the optimization space and criterion discussed in \cite{DENGH1996Sobs} are closely related to ours, we follow a similar approach in designing our simulated annealing algorithm with only a few modifications.

The first modification we make to the simulated annealing algorithm of \cite{DENGH1996Sobs} is to increase the number of reachable vectors in the state update step of the algorithm. This is done by flipping the sign of up to $W$ spreading gains in a single iteration instead of flipping just one, as is done in \cite{DENGH1996Sobs}. At each iteration, the number of spreading gains to change is chosen from a uniform distribution, $X \sim {\cal U}(1,W)$. Then, $X$ randomly selected spreading gains are changed to update the state vector. We found that a modest increase of $W$ from 1 to 2 leads to a lower cost function, hence, a better design.

The second modification we propose is the use of different cost functions than the aperiodic autocorrelation used in \cite{DENGH1996Sobs}. Through experimental studies, we have realized that the choice of a good cost function depends on the system implementation. Here, we are interested in two implementations of OFMT-SS system.  The first implementation emphasizes on keeping the transmit power at a minimum value. For this case, a single spreading code should be used, allowing transmission of a single bit (using a binary shift keying (BPSK) data symbol) or a pair of bits (using a quadrature phase shift keying (QPSK) data symbol) per code interval. For the second implementation, the spreading gains $\bar{\bm\gamma}$ are chosen in each code interval from a set of $M$ multi-codes. This, as discussed below, allows transmission of more information bits per code interval. These two implementations and the relevant cost functions/code designs are discussed in the sequel.

\subsection{Design for single codes} \label{ssec:single_code_papr}
When designing a spreading coefficient vector for use with a single transmission code, we choose the crest-factor as the cost function for the simulated annealing algorithm. We found that directly minimizing the crest-factor leads to a lower PAPR than what we would obtain using the aperiodic autocorrelation cost function proposed in \cite{DENGH1996Sobs}. The PAPR performance of the codes generated with this method are summarized in Table~\ref{tab:PAPR_comparisons1} and compared with the PAPR performance of an FMT-SS operating with optimized spreading gains designed using the method discussed in \cite{LarawayStephenAndrew2015Hbfb}. For the cases presented here, the number of subcarriers in FMT-SS is set to $N=64$ and for OFMT-SS, we let $L=2N=128$. As seen, despite the constraints on the permissible choices of the coefficients $\bar\gamma_k$, the PAPR loss remains relatively low; only $5.07-3.61=1.46$~dB. We can also see that our simulated annealing approach generates spreading gain sets with a crest-factor significantly lower than 2, reported in \cite{1085837}.

\begin{table}
  \caption{An example of PAPR results, comparing OFMT-SS with FMT-SS\label{tab:PAPR_comparisons1}}
  \centering
  \begin{tabular}{|c || c | c |}
    \hline
    Spreading Gain Set & Crest Factor & PAPR (dB) \\ [0.5ex]
    \hline \hline
    FMT-SS             & 1.07         & 3.61      \\
    \hline
    OFMT-SS            & 1.43         & 5.07      \\
    \hline
  \end{tabular}
\end{table}

\subsection{Design for multi-codes} \label{ssec:multi_code_design}
Application of multi-codes to FMT-SS is presented in \cite{HaabDavidB2018FBMS}. Here a trivial method of finding a set of orthogonal codes that minimally affect the PAPR of the optimized design for a single code was proposed. In this work, biorthogonal signaling was used to increase the data rate---i.e. spreading gain vectors were selected from a set of vectors $\Gamma = \left\{\pm \bm{\gamma}_0, \pm \bm{\gamma}_1, \ldots, \pm \bm{\gamma}_{M-1}  \right\}$ where $\bm \gamma_i^\T \bm \gamma_j = 0 \text{ for } i \ne j$. Because positive and negative versions of each spreading gain vector are included in $\Gamma$, it has a cardinality of $2M$ vectors, and, thus,  $B=\log_2(M)+ 1$ bits are transmitted per code interval, if all the codes are deployed. Noting that the number of code vectors can be as large as the code length, OFMT-SS allows doubling the number of  code vectors when compared to FMT-SS.

For FMT-SS, a set of orthogonal multi-code vectors can be generated from an optimized spreading gain vector $\bm\gamma_0$ by point-wise multiplying $\bm\gamma_0$ by the columns of the DFT matrix of the same size \cite{HaabDavidB2018FBMS}. In this case each multi-code vector has the same crest-factor as the original optimized vector. Point-wise multiplying by the columns of the DFT matrix modulates the spreading gain vector with complex sinusoids of different frequencies, which is a linear phase shift of the coefficients. The crest-factor remains unchanged because the spreading gain coefficients are the discrete Fourier series (DFS) coefficients of the time domain signal $m(t)$ and applying a linear phase shift to the coefficients results in a time shift of $m(t)$. This clearly does not change the crest-factor of $m(t)$, hence, has little effect on the PAPR \cite{HaabDavidB2018FBMS}.

Unfortunately, the method proposed in \cite{HaabDavidB2018FBMS} cannot be applied to OFMT-SS. As noted before, while in FMT-SS the spreading coefficients belong to a continuous set, in OFMT-SS we are limited to a discrete set, which cannot be generated by point-wise multiplying by the columns of the DFT matrix. Instead of using the DFT matrix, an alternative approach is to point-wise multiply $\bm{\zeta}_0$ by the columns of a Hadamard matrix. This generates an orthogonal set of multi-code vectors based on an optimized spreading gain vector. Unfortunately, only the first two columns of the Hadamard transform introduce a linear phase shift to $\bm{\zeta}_0$. These two columns result in a good PAPR performance, but the other multi-code vectors result in much higher PAPR. This can be seen in the first row of Table~\ref{tab:PAPR_comparisons2}. When 1 or 2 multi-code vectors are used, the PAPR of the Hadamard multi-coded waveform remains low, but as the number of multi-codes increases, the PAPR increases dramatically. For this reason, we propose an alternative approach for the cases where a larger number of multi-codes should be used.

To come up with a manageable design, we start with a proper choice of the vector $\bm \zeta$, say, ${\bm \zeta}_0$, and use this choice and its circularly shifted versions as a set of codes for our multi-code signaling. But, given the fact that the elements of $\bm\zeta$ are limited to the choices of $+1$ and $-1$, one may note that there is no choice of ${\bm \zeta}_0$ that can lead to a set of perfectly orthogonal codes. We thus argue that a reasonably good choice of ${\bm \zeta}_0$ may be the one that minimizes the correlations between ${\bm\zeta}_0$ and its circularly shifted versions.  With this argument, for our design here, we set the cost function in the simulated annealing algorithm to be the sum of the magnitude squares of the non-zero lag cyclic autocorrelation coefficients of ${\bm\zeta}_0$:
\begin{equation}
  J = \sum_{\ell=1}^{L-1}|\bm{\zeta}_0^\T \bm{\zeta}_\ell|^2
\end{equation}
where
\begin{equation}\label{eq:spreading_gain_vector_cshift}
  \bm{\zeta}_\ell = \begin{bmatrix}
    \zeta_\ell & \zeta_{\ell+1} & \cdots & \zeta_{L-1} & \zeta_0 & \zeta_1 & \cdots & \zeta_{\ell-1}
  \end{bmatrix}^\T.
\end{equation}
This design leads to a set of spreading gain vectors that are {\em approximately/quasi} orthogonal.

The theoretical error probabilities for this class of non-orthogonal signals is beyond the scope of this paper, but are presented in \cite{non_ortho_paper}, where it is seen that if the correlation between different code vectors is kept small, the SER performance has only a small amount of loss compared to the orthogonal code vector case. Here, the small  performance loss of our design is confirmed through numerical simulations in Section~\ref{sec:simulations}. Other works have had similar findings when using non-orthogonal signaling schemes, e.g. \cite{1057730,6773686,RUGINI2019124,1495013}.

In OFMT-SS, with code length $L$, for a given transmission, the number of multi-codes in use can be any value $M$ in the range $1$ to $L$. If a BPSK symbol is added to modulate each of the individual codes, we will have a biorthogonal signaling, \cite{ProakisJohnG2008Dc}. In this case, the transmission rate will be $B=\log_2 M+1$ bits per code interval.

Construction of a set of quasi orthogonal codes as discussed above offers the following advantages.
\begin{itemize}
  \item
        The PAPR of synthesized multi-code signal remains moderately low and nearly independent of the number of multi-codes in use, $M$. Reasonings are presented in Appendix~\ref{app:multi_code_papr}.
  \item
        It allows a low complexity implementation of the receiver. Details of this implementation is presented in Appendix~\ref{app:multi_code_imp}.
\end{itemize}

\begin{table*}[t]
  \centering
  \caption{PAPR results of multi-code signals for Hadamard-based and autocorrelation-based designs.\label{tab:PAPR_comparisons2}}
  \begin{tabular}{|>{\centering\arraybackslash}m{1in} || >{\centering\arraybackslash}m{0.45in} | >{\centering\arraybackslash}m{0.45in} | >{\centering\arraybackslash}m{0.45in} | >{\centering\arraybackslash}m{0.45in} | >{\centering\arraybackslash}m{0.45in} | >{\centering\arraybackslash}m{0.45in} | >{\centering\arraybackslash}m{0.45in} | >{\centering\arraybackslash}m{0.45in} |  }
    \hline
                          & \multicolumn{8}{c|}{$B$}                                                    \\
    \cline{2-9}
                          & 1                        & 2    & 3    & 4    & 5    & 6    & 7     & 8     \\
    \hline \hline
    Hadamard-based        & 5.07                     & 5.08 & 8.31 & 9.32 & 9.46 & 9.55 & 11.16 & 11.13 \\
    \hline
    Autocorrelation-based & 7.53                     & 7.54 & 8.09 & 8.10 & 8.54 & 8.53 & 8.45  & 8.41  \\
    \hline
  \end{tabular}
\end{table*}

To appreciate the multi-code design proposed here, we explore the PAPR values that it delivers. In our study, we used an OFMT-SS system designed with $L=128$ subcarriers. Table~\ref{tab:PAPR_comparisons2} compares both spreading gain design approaches discussed in this paper. As seen, our design for a single code may be preferred only when the number of multi-codes is relatively small; four or smaller. Note that four codes translates to $B=\log_2 4 +1=3$ bits per code interval. For a larger number of multi-codes, our second design performs significantly better. It should be noted that there is a small increase in PAPR as the number of codes grows. This is caused by the interaction of the codes in adjacent symbols.

\begin{figure}
  \centering
  \includegraphics[width=0.95\columnwidth]{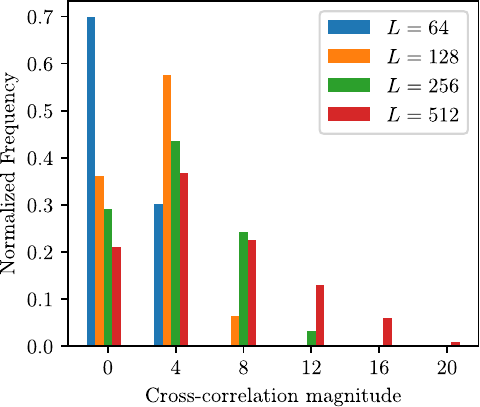}
  \caption{Histograms of cross-correlations of the multi-codes of different lengths $L$.}
  \label{fig:cross_corr}
\end{figure}

Fig.~\ref{fig:cross_corr} presents a histogram of the cross-correlations of a number of choices of multi-codes with different lengths, $L$. As seen, for the case $L=64$, there are two values of cross-correlations that are observed, $0$ or $4$. For the case $L=128$, the observed cross-correlation values are $0$, $4$, or $8$. And, similarly for the rest. The fact that the cross-correlations are all multiples of $4$ relates to the specific choices of $L$ values, here. A derivation showing how these values relate to $L$ is presented in Appendix~\ref{appdx:xcorrs}. In particular, when $L$ is a multiple of $4$, the cross-correlations can take values that are only multiples of $4$, matching our observation in Fig.~\ref{fig:cross_corr}.

To quantify the cross-correlation values in Fig.~\ref{fig:cross_corr}, they should be compared against the magnitude square of the respective code set. Recalling \eqref{eq:no_ici_full_samp_rate_constraint}, the magnitude square that we are referring  to here has the value of
\begin{equation}
  \bm{\bar\gamma}^{\rm H}\bm{\bar\gamma}=L.
\end{equation}
For these comparisons, we may equivalently look at the normalized values of the cross-correlations.
For instance, from Fig.~\ref{fig:cross_corr}, we observe that in the case where $L=64$, about 70\% of the codes have a normalized cross-correlation of $0$ (i.e., they are perfectly orthogonal) and the remaining codes have a normalized cross-correlation of $\frac{4}{64}=0.0625$, which indicates these remaining codes are nearly orthogonal. Similar workouts on the results in Fig.~\ref{fig:cross_corr} reveal that the multi-code design that is proposed in this paper leads to nearly orthogonal codes with the normalized cross-correlations reducing as the code length $L$ increases. The theoretical analysis presented in \cite{non_ortho_paper} reveals that this level of the orthogonality loss has a minimal impact on the system performance. This point is also confirmed through our simulation results in Section~\ref{sec:simulations}.

\section{Clipping for PAPR Reduction} \label{sec:papr_reduction}
Many PAPR reduction methods, mostly, intended for use with OFDM signals have been suggested in the literature; see \cite{4446229} for a comprehensive review. These methods include code and spreading gain design \cite{796380,1085837}, clipping \cite{SDRB,5628256,6469001}, non-linear companding  \cite{4251141}, and tone reservation (TR) and tone injection (TI)\cite{7271004,1233010}, among others. The circularly shifted multi-coding method discussed in the previous section may be classified as a spreading gain design method. Among the other PAPR reduction methods, the non-linear companding method or clipping method may be used to further reduce the PAPR beyond the result obtained through the spreading gain design. Though non-linear companding has been shown to have exceptional PAPR reduction performance, it requires that the channel equalization be applied at the receiver before removing the companding effect \cite{4446229,4251141}. This, besides increasing the complexity of the receiver, removes some of the desirable properties of OFMT-SS that lead to robust performance in harsh environments. An example of such properties is the use of normalized matched filter which blindly removes strong partial-band interferers, \cite{Taylor_NMF}.

Clipping, as the name implies, removes the synthesized signal peaks that are above  a certain threshold to reduce PAPR. If the signal is directly clipped to a threshold, a significant out-of-band energy may be generated. To limit the out-of-band energy, clipping can be done by adding a passband pulse that covers the band of transmission, but has a negated amplitude of each signal peak. This assures a minimum out-of-band energy \cite{SDRB,6469001}. Alternatively, the signal can be repeatedly clipped and filtered to minimize the out-of-band energy \cite{5628256}.

Clipping, naturally, introduces some distortion and, as a result, some degradation in performance should be expected. In OFDM, to keep such distortion minimal, especially when large symbol constellations (like 64-QAM or larger) are in use, only a small PAPR reduction (in the order of 1 or 2 dB) may be possible. This is significantly different in OFMT-SS where symbol constellations are limited to BPSK and QPSK, and multi-coding is used to increase the number of transmitted bits per code interval. In addition, the despreading at the receiver removes most of the induced in-band clipping noise. As a result, clipping allows a significant reduction of PAPR in OFMT-SS with minimal impact on its performance. The numerical results presented in the later parts of this paper confirm this point.

\subsection{Error Vector Magnitude}
In the context of PAPR, in OFDM, the error vector magnitude (EVM) is defined as the root-mean-square (RMS) of the error between the distorted symbols resulting from clipping effect and the original symbols, \cite{6469001, 5628256}. Here, we study the EVM of the individual chips at OFMT-SS subcarriers, as well as the EVM after correlating the received signal with the multi-code spreading gain vectors, i.e., after despreading. By evaluating the EVM before and after correlation, one can compare the OFMT-SS performance with OFDM in the context of PAPR reduction and show the robustness of the former.

Let $\zeta_k[n] = -j^k \gamma_k [n]$ be the $k^{th}$ chip of the $n^{th}$ information symbol, where $\gamma_k[n]$ is the corresponding spreading gain. Here $\zeta_k[n] \in \left\{-1, 1 \right\}$. After applying a clipping noise $\nu(t)$ to the transmit signal, the received chip, discounting channel noise, is
\begin{equation}
  \hat\zeta_k[n] = \zeta_k[n] + \nu_k[n],
\end{equation}
where
\begin{equation}
  \nu_k[n] = h_k(t) \star \nu(t) \biggr\vert_{t=nT + \tau},
\end{equation}
and $\tau$ is the symbol timing offset. Considering the fact that  the transmit signal is a zero mean random process, one finds that $\E\left[ \nu_k[n] \right] = 0$, where $\E[\cdot]$ refers to the statistical expectation.
Next, to simplify the equations that follow, we remove the symbol index from the involved variables.

We define the EVM at chip-level of OFMT-SS as
\begin{align}\label{eq:chip_evm}
  \text{EVM}_{\rm chip} & = \sqrt{\frac{1}{L}\sum_{k=0}^{L-1}\E\left[ \left(\zeta_k - \hat\zeta_k \right)^2  \right]} \nonumber \\
                        & = \sqrt{\frac{1}{L}\sum_{k=0}^{L-1}\Var\left[ \nu_k \right]},
\end{align}
This equation follows the definition of EVM in OFDM, \cite{6469001,7271004}, where the data symbols are equivalent to chips in OFMT-SS.

The symbol level EVM of OFMT-SS, on the other hand, is calculated after correlating the received chips with the associated multi-code vector. The result is
\begin{align}
  \text{EVM}_{\rm symb} & = \sqrt{\E\left[ \left( \frac{1}{L} \left( \bm \zeta^\T \bm \zeta - \bm \zeta^\T  \hat{\bm{\zeta}}] \right) \right)^2 \right]} \nonumber \\
                        & = \sqrt{\E\left[ \left(\frac{1}{L} \bm \zeta^\T \bm \nu  \right)^2 \right]}
\end{align}
where
\begin{equation}
  \bm \nu  = \begin{bmatrix}
    \nu_0 & \nu_1 & \cdots & \nu_{L-1}
  \end{bmatrix}^\T.
\end{equation}

Recalling that $\E\left[\bm\nu \right] = 0$ and only adjacent elements of $\bm \nu$ correlate, thanks to the filtering structure of OFMT-SS, one finds that, \cite{Farhang-BoroujenyB2013Afta},
\begin{align}\label{eq:dspr_evm}
  \text{EVM}_{\rm symb} & = \sqrt{\frac{1}{L^2}\Var\left[ \sum_{k=0}^{L-1}\zeta_k \nu_k\right]} \nonumber                                                               \\
                        & = \frac{1}{L} \sqrt{\sum_{k=0}^{L-1}\Var\left[ \nu_k\right] + 2 \sum_{k=0}^{L-2}\Cov\left[\zeta_k\nu_k, \zeta_{k+1}\nu_{k+1}\right]}\nonumber \\
                        & \approx\frac{1}{L} \sqrt{\sum_{k=0}^{L-1}\Var\left[ \nu_k\right]}
\end{align}
where the approximation in the third line follows since the second summation in the second line is significantly smaller than the first summation. Numerical results that confirm this approximation will be presented later.

By comparing \eqref{eq:chip_evm} and \eqref{eq:dspr_evm}, one finds that  despreading results in a signal to interference ratio (SIR) gain of $10\log L$~dB. Here, interference comes from the clipping. A lower EVM, obviously, allows the use of lower clipping thresholds with a very limited effect on the $\text{EVM}_{\rm symb}$ and, as a result, BER performance. It thus makes clipping an excellent candidate for further reduction of the PAPR in OFMT-SS.

Next, we study the signal to interference plus noise ratio (SINR), where the interference is the clipping noise, and noise refers to channel additive noise. To determine the SINR, without any loss of generality, we assume that the transmitted symbol (a binary bit taking values $\pm 1$) is $+1$. Also, for clarity of our derivations, we assume that at the detector output bits are scaled by $\sqrt{\En}$, hence, bit (signal) power is equal to $\En$. Accordingly, the bit estimate is obtained as
\begin{align} \label{eq:rx_clip_plus_noise}
  \hat{s} & = \frac{1}{L}\sum^{L-1}_{k=0}\left(\sqrt{\En} \left(\zeta_k+ \nu_k \right) + w_k \right) \zeta_k     \nonumber \\
          & = \sqrt{\En} + \frac{\sqrt{\En}}{L} \sum_{k=0}^{L-1}\nu_k \zeta_k + \frac{1}{L}\sum_{k=0}^{L-1}w_k \zeta_k.
\end{align}
In the right-hand side of \eqref{eq:rx_clip_plus_noise}, the signal portion is the first term, the interference is the second term, and the channel noise is the third term.

Appendix~\ref{appndx:noise correlation} presents a statistical analysis of the noise terms $w_k$. It is shown that the noise terms $w_k$ are a set of zero-mean independent and identically distributed (iid) random variables, with the autocorrelation/covariance matrix \begin{equation}\label{eq:Cww}
  {\bf C_{ww}} =\frac{\sigma^2}{2}\textbf{I}
\end{equation}
where $\sigma^2$ is the one-side power spectral density of the channel noise.

An analysis of the clipping noise terms $\nu_k$ turns out to be difficult. Here, we take note that for most cases of interest, the clipping noise has a negligible impact on the receiver performance. With this point in mind, to allow the following derivations and analysis, we simply assume the clipping noise samples $\nu_k$ are also zero-mean and iid and have the covariance matrix
\begin{equation}\label{eq:Cvv}
  {\bf C}_{\bm \nu\bm \nu} =\text{EVM}_{\rm symb}^2\textbf{I}.
\end{equation}
Moreover, we assume that the channel noise samples $w_k$ and the clipping noise samples $\nu_k$ are two independent sets.

Making use of \eqref{eq:Cww} and \eqref{eq:Cvv} and taking note that $\zeta_k\in [-1,1]$, it is straightforward to show that
\begin{equation}
  \sigma_{\rm c}^2 = \En\text{EVM}_{\rm symb}^2
\end{equation}
and
\begin{equation}
  \sigma_{\rm n}^2 = \frac{\sigma_w^2}{L},
\end{equation}
where $\sigma_{\rm c}^2$ is the variance of the second term in the right-hand side of \eqref{eq:rx_clip_plus_noise}, i.e., the clipping noise term, and $\sigma_{\rm n}^2$ is the variance of the third term in the right-hand side of \eqref{eq:rx_clip_plus_noise}, i.e., the channel noise term.

Making use of the above results, we get
\begin{align}\label{eq:SINR}
  \text{SINR} & = \frac{\En}{\sigma_c^2 + \frac{\sigma_w^2}{L}}    \nonumber              \\
              & = \frac{1}{\text{SNR}\times\text{EVM}_{\rm symb}^2 + 1}\times \text{SNR},
\end{align}
where
\begin{equation}
  \text{SNR} = L\frac{2\En}{\sigma^2}
\end{equation}
is the $\text{SNR}$ at the receiver output. The numerical results presented later reveal that, in typical applications of OFMT-SS, the SNR degradation resulting from clipping remains negligible, i.e., $\text{SNR}\times\text{EVM}_{\rm symb}^2 \ll1$, hence, $\text{SINR}\approx \text{SNR}$.

To quantify our theoretical result in \eqref{eq:SINR}, Table~\ref{tab:EVM_cmp} presents a set of results showing how the choice of clipping threshold impacts $\mbox{EVM}_{\rm chip}$ and $\mbox{EVM}_{\rm symb}$. These results correspond to the case where a biorthogonal signaling is used and $L=128$. A clipping algorithm similar to those presented in \cite{SDRB,5628256,6469001} is applied for different choices of the clipping threshold. As one would expect, both EVMs increase as the clipping threshold decreases. Nevertheless, even for a threshold as low as $3.75$~dB, $\mbox{EVM}_{\rm symb}$ remains very small. If one substitutes such values and a typical SNR value in the range of $10$ to $20$~dB in \eqref{eq:SINR}, he will find the SNR loss arising from PAPR reduction translates to very small value; less than $0.1$~dB. The SER results presented in the next section confirm this negligible loss. It is also worth comparing the results in the second and third columns of Table~\ref{tab:EVM_cmp}, and note that  $\mbox{EVM}_{\rm symb}=\mbox{EVM}_{\rm chip}/\sqrt{L}$. Taking note that $L$ is the receiver processing gain, this shows, it is the despreading effect at the receiver that reduces the $\mbox{EVM}_{\rm symb}$ to such a low values.

\begin{table}[t]
  \caption{\label{tab:EVM_cmp}$\mbox{EVM}_{\rm chip}$ and $\mbox{EVM}_{\rm symb}$ results for different clipping thresholds $\mu$.}
  \centering
  \begin{tabular}{|c ||c | c |c |}
    \hline
    Threshold (dB) & $\mbox{EVM}_{\rm chip}$ & $\mbox{EVM}_{\rm symb}$ \\
    \hline \hline
    3.75           & 0.1174                  & 0.0094                  \\
    \hline
    4.00           & 0.1034                  & 0.0085                  \\
    \hline
    4.25           & 0.0903                  & 0.0075                  \\
    \hline
    4.5            & 0.0775                  & 0.0066                  \\
    \hline
    4.75           & 0.0655                  & 0.0057                  \\
    \hline
    5.00           & 0.0548                  & 0.0049                  \\
    \hline
  \end{tabular}
\end{table}

\section{BER and SER results} \label{sec:simulations}
In this section, we extend the study of OFMT-SS waveform by looking at a few examples of its bit error rate (BER) and symbol error rate (SER). In each case, we use an OFMT-SS signal with $L=128$. The code designs for single-code and multi-codes follow those presented in Section~\ref{sec:ofmt_papr}.

\subsection{Single-code Performance}
\begin{figure}
  \centering
  \includegraphics[width=0.95\columnwidth]{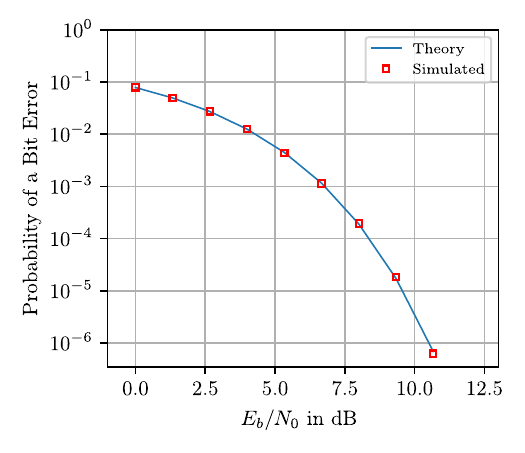}
  \caption{Simulated BER curve of single code OFMT-SS. The QPSK theoretical BER curve is presented for comparison.}
  \label{fig:uncoded_ber_curve}
\end{figure}

For the single code case, we use the method discussed in Section~\ref{ssec:single_code_papr} to design a spreading gain set. The BER of this system transmitting quadrature phase shift keying (QPSK) information symbols is plotted against $E_b/N_0$ in Fig.~\ref{fig:uncoded_ber_curve}. This figure includes the theoretical QPSK curve, showing a perfect match with our simulated results. This demonstrates the Nyquist, i.e., no inter-symbol interference (ISI), property of the OFMT-SS waveform. The receiver here is a matched filter, followed by a sampler that takes signal samples at the peak of the system impulse response.

\subsection{Multi-code Performance} \label{ssec:sim:multi-code}
For the multi-coded case, we evaluate the symbol error rate (SER) instead of the bit error rate, which allows for easy comparison with the SER probabilities given in \cite{ProakisJohnG2008Dc}.

\begin{figure}
  \centering
  \includegraphics[width=\columnwidth]{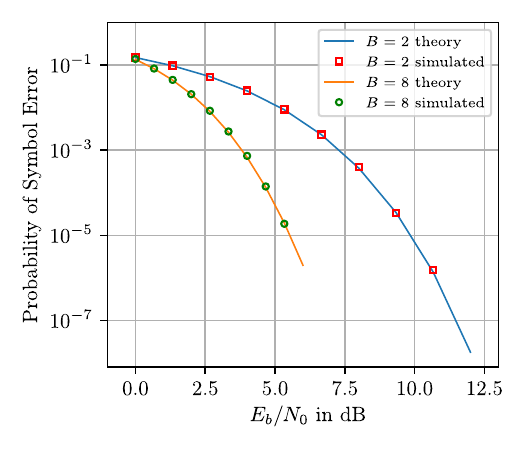}
  \caption{Simulated SER curves multi-code OFMT-SS for the cases of $B=2$ and $B=8$ bits per symbol. Theoretical curves  from \cite{ProakisJohnG2008Dc} are presented for comparison.}
  \label{fig:ser_no_reduction}
\end{figure}

Using a system with $L=128$ multi-code vectors, we can achieve a maximum bit rate of $\log_2(L) + 1=8$ bits per code interval.  As discussed in \cite{ProakisJohnG2008Dc}, one may choose to use a subset of the multi-codes, resulting in a smaller number of bits, $B$, per code interval. Fig.~\ref{fig:ser_no_reduction} shows the simulated SER for the cases of $B=8$ and $B=3$ along with the corresponding theoretical orthogonal theory curves. From this figure, we can see that although, here, the code vectors are not perfectly orthogonal, the loss between the orthogonal theory curve and the simulated SER, using quasi orthogonal code vectors, is very small/unnoticeable.  This observation is in line with our theoretical findings in \cite{non_ortho_paper}.

\subsection{PAPR Reduction Impact} \label{ssec:sim:peak_cancel}
\begin{figure}
  \centering
  \includegraphics[width=\columnwidth]{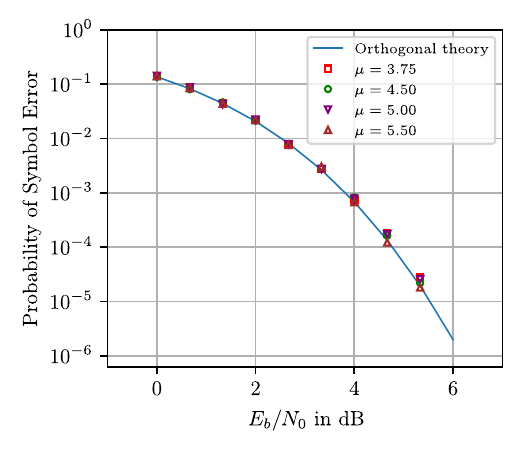}
  \caption{Simulated SER curves for several PAPR reduction thresholds $\mu$. The results are for the case where multi-codes and biorthogonal signaling are deployed, and $B=8$ bits are transmitted per code interval. Other parameters: $L=128$, $K=64$, $P=2$. Hard clipping is applied after the second iteration of the clipping algorithm.}
  \label{fig:papr_reduction_ser_curve}
\end{figure}

To see the impact of the clipping on the SER performance, we consider a number of clipping thresholds and compare their performance against the theoretical biorthogonal signaling theory curve. Fig.~\ref{fig:papr_reduction_ser_curve} presents the SER curves for several values of the clipping threshold, $\mu$. The results correspond to the case where $L=128$,  and $B=8$ bits are transmitted per code interval. From these results, we can see that there is only a very small loss introduced by clipping, confirming our findings in Section~\ref{sec:papr_reduction}.

\section{Conclusion}\label{sec:conclusion}
We introduced and explored the details of a new filter bank-based waveform for spread spectrum communications. The proposed waveform is a modified form of filtered multi-tone spread spectrum (FMT-SS). We took note that in a conventional FMT-SS, subcarrier bands are non-overlapping. This results in a clear foot-print  in the FMT-SS signal spectra  that may reveal some of its characteristics to an unauthorized receiver. The new waveform, which we refer to as overlapped FMT-SS (OFMT-SS), adds additional subcarriers to fill in the empty spectra between the adjacent subcarrier bands of FMT-SS, leading to a flat spectrum. We studied the choices of the spreading gains in OFMT-SS and found the constraints that should be imposed to these gains to result in a perfectly flat synthesized transmit signal. We also developed a method of optimizing the spreading gains to minimize the PAPR of the synthesized signal. Moreover, to further reduce the PAPR, we explored the use of signal clipping methods and showed that in the particular case of OFMT-SS, PAPR values of as low as $4$~dB or even smaller, without much impact on the system performance, was possible.

\section*{Acknowledgements}
\thanks{This manuscript has in part been authored by Battelle Energy Alliance, LLC under Contract No. DE-AC07-05ID14517 with the U.S. Department of Energy. The United States Government retains and the publisher, by accepting the paper for publication, acknowledges that the United States Government retains a nonexclusive, paid-up, irrevocable, world-wide license to publish or reproduce the published form of this manuscript, or allow others to do so, for United States Government purposes. STI Number: INL/JOU-25-83708.}

\appendices

\section{PAPR Invariance of Multi-codes} \label{app:multi_code_papr}
In this appendix, we show that the PAPR minimally varies between different choices of the multi-codes when such codes are being generated according to the method discussed in Section~\ref{ssec:multi_code_design}. To this end, we first recall that there is a close relationship between the crest-factor of $\bar m(t)$ and the PAPR of the synthesized signal; e.g, see \cite{LarawayStephenAndrew2015Hbfb}. Then, we concentrate on the crest-factor $\bar m(t)$ and show that it is invariant of the choices of the multi-code vectors when they are selected according to \eqref{eq:spreading_gain_vector_cshift}.

A proof to the latter statement can be made by first noting that
\begin{equation}\label{eq:m'(t)}
  \bar m'(t)=\sum_{k=0}^{L-1} \zeta_k e^{j2\pi f_k t}
\end{equation}
has the same crest-factor as $\bar m(t)$. This follows if one takes note that $\bar m(t)$ is obtained from $\bar m'(t)$ by adding a linear phase to the coefficients $\zeta_k$, namely, the addition of the $j^k$ coefficients in \eqref{eq:no_ici_full_samp_rate_constraint}. Such linear phase results in a signal shift which clearly has no impact on its crest-factor. On the other hand, taking note that \eqref{eq:m'(t)} is an inverse Fourier series, a circular shift of its coefficients translates to a modulation, i.e., multiplication of the synthesized signal by an exponential factor $e^{j2\pi f_c t}$, where $f_c$ is the modulation frequency. This, also, does not change the crest-factor of the resulting signal.

\section{Low Complexity Implementation of the Receiver} \label{app:multi_code_imp}
A generic implementation of the receiver involves correlation of the signal samples at the output of the analysis filter bank (after equalization) with each of the multi-codes separately. For a code length $L$ and when the maximum number of multi-codes, equal to $L$, are used, completing all the correlations requires $L^2$ multiply and add operations. The choice of multi-codes according to \eqref{eq:spreading_gain_vector_cshift} allows one to reduce this complexity to about $L\log_2 L$ operations through the use of fast Fourier transforms (FFTs). Next, for completeness of our discussion, we present details of this correlator.

Let $\bf J$ be an  $L\times L$ circulant matrix with the first column ${\bm\zeta}_0$ and $\bf y$ denote the vector of outputs of the analysis filter bank, after equalization. The vector of correlations of interest is given by
\begin{equation}\label{eq:c vector}
  {\bf c}={\bf J}^{\rm H}({\bf b^*}\odot{\bf y}).
\end{equation}
Next, we note that since $\bf J$ is a circulant matrix, it may be expanded as, \cite{GrayRobertM2006Tacm},
\begin{equation}\label{eq:Jfactorization}
  {\bf J}=\Fourier^{\rm H}{\bm \Lambda}\Fourier
\end{equation}
where $\bm \Lambda$ is a diagonal matrix whose diagonal elements are obtained by taking the discrete Fourier transform (DFT) of $\bm\zeta_0$ and $\Fourier$ is the normalized DFT matrix, satisfying $\Fourier^{\rm H}\Fourier={\bf I}$.

Substituting \eqref{eq:Jfactorization} in \eqref{eq:c vector}, we get
\begin{equation}\label{eq:c vector2}
  {\bf c}=\Fourier^{\rm H}{\bm \Lambda}^*\Fourier({\bf b^*}\odot{\bf y}).
\end{equation}
Implementation of \eqref{eq:c vector2} involves $L$ multiplications, to perform $({\bf b}\odot{\bf y})$, $L\log_2 L$  to perform multiplications with $\Fourier$ and $\Fourier^{\rm H}$, and $L$ multiplications to include ${\bm \Lambda}^*$ in the operations.

\section{Noise Correlation Properties at\\ the Despreader Input}\label{appndx:noise correlation}

Consider a multi-coded OFMT-SS system where the spreading gain vector at the $n^{th}$ code interval is
\begin{equation}
  \bm \gamma[n] = \begin{bmatrix}
    \gamma_0[n] & \gamma_1[n] & \cdots & \gamma_{L-1}[n]
  \end{bmatrix}^\T.
\end{equation}

The signal at the $k^{th}$ analysis filter bank output is given by
\begin{align} \label{eq:y(t)}
  y_k(t) = (-j)^k & h_k^*(-t) \star    \left(\sum_{n=-\infty}^{\infty} \Bigl[ \gamma_k[n]h_k(t - nT)\bigr.\right.      ~~~~~~~~~~~~~                           \nonumber \\
                  & ~~~~~~~~~~~~~~~~~~~~~~ + \gamma_{k-1}[n]h_{k-1}(t-nT) \nonumber                                                                                      \\
                  & ~~~~~~~~~~~~~~~~~~~~~~ + \gamma_{k+1}[n]h_{k+1}(t-nT) \Bigl. \Bigr] \nonumber                                                                        \\
                  & ~~~~~~~~~~~~~~~~~~~~~~ + \tilde{w}(t)\bigl. \biggr),
\end{align}
where
\begin{align}
  h_k(t) = h(t)e^{j2\pi f_k t},
\end{align}
$\tilde{w}(t)$ is a zero mean, complex white Gaussian random process with one-sided power spectral density  $\sigma^2$, and we define $\gamma_{-1} = \gamma_L = 0$.

By performing straightforward manipulations and substituting $\gamma_k[n] = j^k\zeta_k[n]$, \eqref{eq:y(t)} can be rearranged as
\begin{align} \label{eq:y_s1(t)}
  y_k(t) = & \sum_{m=-\infty}^{\infty} \zeta_k[m] \rho(t-mT)\nonumber                               \\
           & ~~~~~~~~~~~~~~~ - j \sum_{m=-\infty}^{\infty}\zeta_{k-1}[m]d_{k-1}^*(-(t-mT))\nonumber \\
           & ~~~~~~~~~~~~~~~ + j \sum_{m=-\infty}^{\infty} \zeta_{k+1}[m]d_k(t-mT)        \nonumber \\
           & ~~~~~~~~~~~~~~~ + \tilde{w}_k(t),
\end{align}
where $\rho(t)$ is given in \eqref{eq:fmt_rho}, $d_k(t) = d(t)e^{j2\pi f_k t}$, $d(t)$ is given in \eqref{eq:single_ici_term}, and $\tilde{w}_k(t) = (-j)^k h_k^*(-t) \star \tilde{w}(t)$.

Next, using  \eqref{eq:single_ici_term}, one finds that
\begin{equation}
  D(f) = H(f)H(f-\frac{1}{T}),
\end{equation}
where $H(f)$ and $D(f)$ are the Fourier transforms of $h(t)$ and $d(t)$, respectively.  We also define $D'(f) = D\left(f + \frac{1}{2T}\right)$, and take note that converting it to the time domain leads to
\begin{equation}\label{eq:d(t)}
  d'(t) = d(t)e^{-j\frac{2 \pi t}{2T}}.
\end{equation}
Moreover, using the definition of $\bar f_k$ from \eqref{eq:fkOFMT-SS}, one finds that
\begin{equation}\label{eq:dk(t)}
  d_k(t) = d'(t)e^{j\frac{2\pi}{T}(k-N+1)t}.
\end{equation}

To proceed, we assume that $h(t)$ is chosen to be real-valued and symmetric around $t=0$, i.e., a zero-phase filter. When this condition holds, it is not difficult to show that $d'(t)$ is also a zero-phase pulse and, thus, it is real-valued and symmetric around $t=0$. Also, \eqref{eq:d(t)} and \eqref{eq:dk(t)} imply that the samples
\begin{equation}
  d_k(nT) = d'(nT)=d(nT)
\end{equation}
are a set of real-valued numbers. Using this result in \eqref{eq:y_s1(t)}, sampling at $t=nT$, and recalling that $\rho(t)$ is a Nyquist pulse,
\begin{align}\label{eq:y[n]}
  y_k[n] =  \zeta_k[n] & +j\sum_{m=-\infty}^{\infty}\left( \zeta_{k+1}[m] - \zeta_{k-1}[m] \right)d((n-m)T) \nonumber \\
                       & ~~~~~~~~~~~~~~~~~~~~~~~~~ +\tilde w_k[n],
\end{align}
where $\tilde w_k[n]$ is the sampled complex random variable that originated from the channel noise.

The estimate of $\zeta_k$, denoted by $\hat\zeta_k$, is obtained by taking the real part of $y_k[n]$. Also, taking note that the summation on the left-hand side of \eqref{eq:y[n]} is real-valued, we obtain
\begin{align} \label{eq:zeta_hat}
  \hat{\zeta}_k[n] & = \zeta_k[n] + w_k[n],
\end{align}
where $w_k[n]=\Re\left\{\tilde w_k[n]\right\}$. Assuming the channel noise is a zero-mean process, \eqref{eq:zeta_hat} shows that the $\hat{\zeta}_k[n]$ is an unbiased estimate of $\zeta_k[n]$.

Next, we are interested in autocorrelation $\bf C_{ww}$ of the noise vector
\begin{equation}
  {\bf w}[n]=\left[w_0[n]~w_1[n]~\cdots~w_{L-1}[n]\right].
\end{equation}
To find $\bf C_{ww}$, we begin with the continuous-time vectored random process
\begin{equation}
  \tilde{\bf  w}(t) = \begin{bmatrix}
    \tilde{w}_0(t) & \tilde{w}_1(t) & \cdots & \tilde{w}_{L-1}(t)
  \end{bmatrix}^\T
\end{equation}
and take note that
\begin{equation}
  \tilde{\bf  w}(t)=w(t)\star  {\bf  h}^*(-t),
\end{equation}
where
\begin{equation}
  {\bf h}(t) = \textbf{b} \odot \begin{bmatrix}
    h_0(t) & h_1(t) & \cdots & h_{L-1}(t)
  \end{bmatrix}^\T,
\end{equation}
$\textbf{b}$ is given by \eqref{eq:def_b}, and $\odot$ is an element-wise product. Making use of the results presented in \cite[p. 406-407]{PapoulisAthanasios2002Prva}, the autocorrelation of  $\tilde{\bf w}(t)$ can be obtained as
\begin{equation} \label{eq:cross_corr_def}
  \textbf{R}_{\tilde{\bf w}\tilde{\bf w}}(\tau) = \left(R_{\tilde{w}}(t) \star {\bf h^*}(-t) \star {\bf h}^{\text{T}}(t)\right)\Big|_{t=\tau}
\end{equation}
where $R_{\tilde{w}}(t) = \sigma^2 \delta(t)$ is the autocorrelation function of $\tilde{w}(t)$. Using \eqref{eq:cross_corr_def}, the $k,l^{\text{th}}$ element of $\textbf{R}_{\tilde{\bf w}\tilde{\bf w}}(\tau)$ can be shown to be
\begin{equation} \label{eq:cross_corr_func}
  \left[\textbf{R}_{\tilde{\bf w}\tilde{\bf w}}(\tau)\right]_{k,l} = \begin{cases}
    \sigma^2 \rho(\tau)   & \text{for } k = l     \\
    j \sigma^2 d^*(-\tau) & \text{for } l = k + 1 \\
    -j \sigma^2 d(\tau)   & \text{for } l = k - 1 \\
    0                     & \text{otherwise}
  \end{cases}
\end{equation}

Now, assuming that the channel noise $w(t)$ is a proper (i.e., circularly symmetric) complex Gaussian process, from the results in  \cite[p. 35]{SchreierPeterJ.2010SSPo}, one will find that the vector $\tilde{\bf w}(t)$ is also a proper complex Gaussian process, hence,
\begin{equation} \label{eq:proper_cov_property}
  \textbf{R}_{\bf ww}(\tau) = \frac{1}{2}\Re\left\{ \textbf{R}_{\tilde{\bf w}\tilde{\bf w}}(\tau)\right\}
\end{equation}
where $\textbf{R}_{\bf ww}(\tau)$ is the autocorrelation function of ${\bf w}(t)=\Re\{\tilde{\bf w}(t)\}$. We also note that
\begin{equation}
  {\bf C_{ww}}=\textbf{R}_{\bf ww}(0).
\end{equation}
Taking note that $\rho(0)=1$, $d(0)$ is real-valued, and recalling \eqref{eq:cross_corr_func}, we get equation \eqref{eq:Cww}.

\section{Cross-correlation Terms}\label{appdx:xcorrs}
Consider two spreading chip sequences
\begin{align}
  \bm\zeta^{(0)} = \begin{bmatrix}
                     \zeta_0^{(0)} & \zeta_1^{(0)} & \dots & \zeta_{L-1}^{(0)}
                   \end{bmatrix}^\T \nonumber \\
  \bm\zeta^{(1)} = \begin{bmatrix}
                     \zeta_0^{(1)} & \zeta_1^{(1)} & \dots & \zeta_{L-1}^{(1)}
                   \end{bmatrix}^\T\nonumber
\end{align}
where $\zeta_k^{(l)} \in \left\{ -1, +1 \right\}$ and both $\bm\zeta^{(0)}$ and $\bm\zeta^{(1)}$ contain the same number of $+1$ and $-1$ elements.

The cross-correlation between $\bm\zeta^{(0)}$ and $\bm\zeta^{(1)}$ may be written as
\begin{equation}
  C = \sum_{k=0}^{L-1}x_k
\end{equation}
where $x_k = \zeta_k^{(0)} \zeta_k^{(1)} \in \left\{ -1, +1 \right\}$.

Next, we define the sets $X_p = \left\{ x_k | x_k = +1\right\}$ and $X_n = \left\{ x_k | x_k = -1\right\}$, and the summations $C_p  = \sum X_p$ and $C_n  = \sum X_n$. We take note that $C_p-C_n=L$ and let $C_p+C_n=C$. Using the latter two identities, one finds that
\begin{equation} \label{eq:corr_simp}
  C = L + 2C_n.
\end{equation}
This implies that the integer $C$ has the same parity as $L$.

Now we show that $C_n$ is even. For this, we start with the trivial case
\begin{equation}
  \bm\zeta^{(1)} = \bm\zeta^{(0)},
\end{equation}
where $C_n=0$. Next, $\bm\zeta^{(1)}$ is updated by swapping two of its elements. If these elements have the same sign, there will be no change in the value $C_n$. If they have different sign, $C_n$ will reduce to $-2$, since the product terms, $x_k$, at both positions becomes negative. Repeating this process of swapping elements, $C_n$ only increases or decreases by 2. This implies that $C_n$ always remain a negative even number. Hence, $C$ in \eqref{eq:corr_simp} will take on discrete values in steps of 4. In the case where $L$ is a multiple of $4$, the cross-correlation $C$ can take values $0$, $4$, $8$, $\cdots$. Finally, we argue that this observation is applicable to the construction of the multi-codes discussed in Section~\ref{ssec:multi_code_design}. This is because all circular shifts of $\bm\zeta^{(0)}$ results in vectors in which the number of $+1$ and $-1$ elements are preserved.

\bibliographystyle{IEEEtran}
\bibliography{IEEEabrv,ref, papr_reduction}

\end{document}